\documentclass[12pt,thmsa]{revtex4}
\begin{document}

\title{Normal metal - insulator - superconductor interferometer}

\author{K. Yu. Arutyunov$^{\ast}$ and T. T. Hongisto}
\address {University of Jyv\"askyl\"a, Department of Physics, PB 35,
40014 Jyv\"askyl\"a, Finland}

\date{\today}

\begin{abstract}

Hybrid normal metal - insulator - superconductor microstructures suitable for studying an interference of electrons 
were fabricated. The structures consist of a superconducting loop connected to a normal metal electrode 
through a tunnel barrier . An optical interferometer with a beam splitter can be considered as a classical 
analogue for this system.
All measurements were performed at temperatures well below 1 K. The interference can be observed as periodic 
oscillations of the tunnel current (voltage) through the junction at fixed bias voltage (current) as a function of 
a perpendicular magnetic field. The magnitude of the oscillations depends on 
the bias point. It reaches a maximum at energy $eV$ which is close to the superconducting gap and decreases with 
an increase of temperature. Surprisingly, the period of the oscillations in units of magnetic flux 
$\Delta \Phi$ is equal neither to $h/e$ nor to $h/2e$, but significantly exceeds these values 
for larger loop circumferences. The origin of the phenomena is not clear. 

\end{abstract}

\pacs{7335,7440}

%\begin{keywords}
% nonequilidrium superconductivity, microstructures, Aharonov -Bohm effect 
%\end{keywords}

\maketitle

%  ############### end of front matter###############%

\newpage

The simplest optical interferometer consists of a beam splitter, a pair of mirrors and an opaque screen.
The repositioning of the mirrors causes variation of the light intensity at a given point on the screen. 
In conventional passive media (e.g. air) the wavelength of light is not altered along the optical path. 
A sharp interference pattern requires only a monochromatic source of light. In the same way, a metal loop 
with two electrodes can be considered as an analogue to the described optical interferometer. One node 
corresponds to the electron beam splitter and the second node - to the 'screen'. The total current through 
both paths (at a fixed voltage, for example) is then equivalent to the light intensity. Certainly, one can alter 
the 'interference pattern' in the usual way by changing the loop geometry. However,
there is a more convenient way of doing this: the application of a perpendicular magnetic field. The conductivity 
of a normal metal loop is periodic in units of magnetic flux quantum $\Delta \Phi = \phi _{0}^{N}= h/e$, 
where $e$ is the electron charge \cite{Umbach}. 
Contrary to optics, the phase of the electron wave function inside a normal metal 
interferometer can be randomly altered due to inelastic scattering. Thus, the size of the loop is 
limited by the phase breaking length $\ell _{\varphi}$. Micron-size metal structures at low temperatures 
($T < 1$ K) are subject to this limit \cite {Mesoscopic book}. The utilisation of superconductors should eliminate
this size limitation due to macroscopic quantum coherence. The only problem is that in the pure superconducting 
state, resistive measurements are useless. Other system parameters such as magnetisation or the critical temperature 
\cite {Little-Parks}, \cite {Tinkham book} should be measured.

A number of normal metal-insulator-superconductor (N-I-S) microstructures has been fabricated. They can be considered 
as a solid-state analogue of an optical interferometer with a beam splitter. A closed loop of aluminium is 
overlapped at one point by a copper electrode through a thin oxide barrier (Fig. 1).  
At sufficiently low temperatures, aluminium becomes superconducting. Due to the 
macroscopic phase coherence, there is no random alternation of the electron phase inside the loop of 
a superconducting interferometer, while the finite resistance of the whole N-I-S system enables 
electric measurements. Structures were fabricated by two-angle
metal evaporation through an e-beam patterned double layer P(MMA-MAA)/PMMA mask. Typical aluminium thickness $d_{Al}$ 
was $\sim$ 35 nm and the line width $\sim$ 120 nm for small samples. For larger loops ($>$ 10 $\mu$m) 250 nm lines 
were used. Before deposition of the top copper electrode ($d_{Cu} \sim$ 30 nm) the aluminium surface was oxidised 
in $O_2$ atmosphere at a pressure of about 1 mbar for 1 to 2 minutes. The nominal overlapping area between copper and 
aluminium was about 100 nm $\times$ 100 nm. The tunnel resistance at room temperature varied from sample to 
sample from roughly 1 to 60 $k \Omega$, increasing by $\sim$ 20\%   while cooling down to liquid 
helium temperatures. The majority of experiments were made in a voltage-biased mode (Fig. 1). The derivative of the
variation of the current with respect to the voltage, $dI/dV(V)$ characteristic,
was measured by lock-in technique by using a $\sim$ 1 $\mu$V ac modulation of the biasing voltage.
Current biased dependencies were also studied. The resistance of the metal parts was roughly a few tens of Ohms.
Thus, the dominating part of the voltage drop was due to the tunnel barrier. Experiments were made 
in a $^{3}He-^{4}He$ dilution refrigerator placed inside an electromagnetically shielded room. Only battery powered 
front-end amplifiers were kept inside the room. These were connected to the remaining electronics outside 
through carefully shielded coaxial cables carrying analogue signals. Three stages of filtering 
were used. Firstly, \textit{Spectrum Control} 51-390-305 filters with -80 dB cut-off at 300 kHz were placed on
top of the cryostat at room temperature.  A second stage was located at $\sim$ 1K point consisting of 
capacitive and inductive C-L-C 
elements (220 nF, 2.2 mH, 220 nF) forming a $\pi$-filter. The last stage was made of $\sim$ 30 cm 
\textit{Thermocoax Philips} cable \cite{Zorin} which was thermally anchored to the sample stage. 
The utilisation of filters at room temperature resulted 
in marginal improvement of the signal-to-noise ratio, while the performance of the $\pi$-filter at 1 K
appeared to be crucial since 
no reasonable signal-to-noise ratio could be obtained without it. Magnetic fields up to 30 mT were generated
by various 2 or 4 layer superconducting coils wound directly on outside of the the refrigerator's vacuum canister. 
The coil inputs leading to the current source were also filtered. Magnetic field sweeps (step-by-step) were made 
rather slowly ($\sim$ 2 $\div$ 5 s/point). In some cases a car battery was connected through a decade resistor block
and was 
used as a current source. The Earth's magnetic field was not screened. The latter may explain a small offset of the 
magnetic field  ($\sim$ 0.05 mT) which appears in the data presented below.

At zero magnetic field the current-voltage characteristics, $I(V,B=0)$, show  typical behaviour for N-I-S junctions 
(Fig. 2a). Hereafter the superconducting gap energy $\Delta$ is defined as $\Delta = eV_{gap}$, 
where $V_{gap}$ corresponds to the maximum 
of the $dI/dV(V)$ dependencies. This assumption is quite justified at temperatures well below the critical 
temperature of superconducting electrode when the $dI/dV(V)$ characteristic is 'sharp'. The majority of 
measurements were made at temperatures $<$ 500 mK, while the critical temperature $T_c$ of the co-deposited
aluminium layers was about $\sim$ 1.3 K.

The application of a perpendicular magnetic field modifies the {I(V, B=const)} dependencies in a non-monotonous 
way (Fig. 2b). Sweeping the field at a constant voltage bias demonstrates the nature of the dependency more profoundly.
In this case, the {I(V=const, B)} characteristics are quasi-periodic with respect to the magnetic field (Fig. 3). 
The monotonous envelope 
behaviour can be qualitatively explained by the reduction of the superconducting energy gap by magnetic field. 
The period of oscillations $\Delta B_{I} \equiv \Delta B(V=const, B)$ is not constant, but drops by a few factors 
at high fields 
(Fig. 3, inset). As the last phenomenon is not well understood, hereafter only the low field data 
($\mid B\mid \leq $ 2 mT) is considered, where the magnitude and the period of oscillations are field-independent. 
The $I(V=const, B)$ dependencies are essentially hysteretic (Fig. 4). Phenomenologically one may state that the
$I(V=const, B)$ characteristics form a set of 'parabolas', where allowed current states 'jump' from 
neighbouring branches of parabolas, depending on the direction of the magnetic field sweep. Within the range 
of fields corresponding to a single 'parabola' dependencies are not hysteretic. The $I(V=const, B)$
characteristics are well reproducible and become noisy at biases $V$ noticeably higher than the gap voltage 
$V_{gap}$. The normalised magnitudes of current oscillations $\Delta I/I_{max}$ as function of the 
normalised bias voltage $V / V_{gap}$ for several samples with various loop size are plotted in Fig. 5. 
There are at least two common features. Firstly, the function $\Delta I/I_{max}(V / V_{gap})$ has a maximum 
slightly below the gap voltage $V / Vgap$ $\sim$ 0.7 $\div$ 0.8. Secondly, this maximum is pronounced at lower 
temperatures and becomes smeared at temperatures higher than $\sim$ 500 mK. Surprisingly, the normalized 
magnitude of oscillations $\Delta I/I_{max}$ reaches nearly 100\% at sufficiently low temperatures. 
As a comparison, the Aharonov-Bohm effect in micron-size normal metal rings has 
the magnitude $\Delta R/R < 10^{-3}$ \cite {Umbach}, while in a micron-size superconducting aluminium ring
the magnitude of the critical temperature oscillations (Little-Parks effect) is $\Delta T_c / T_c < 10^{-2}$
\cite {Mosch L-P} , \cite {nonlocality}. The maximum magnitude of the current oscillations decreases slightly with  
an increase of the loop diameter, but the effect is still well pronounced even for a loop size as high as 25 $\mu$m 
(Fig. 7). For small size loops ($<$ 3 $\mu$m) no correlation between the maximum magnitude of current 
oscillations and the tunnel resistance in the range from 3 $k\Omega$ to 60 $k\Omega$ has been found. For loops 
with the side $L > $ 5 $\mu$m no effect has been detected for structures with tunnel resistance $R_T >$ 8 $k\Omega$.

For a given size of loop, the period of current oscillations $\Delta B_{I}$ in low fields depends slightly 
on the bias voltage and increases by $\sim$ 15\% below the gap (Fig. 6 and Fig. 8c). The effect is more pronounced at
low temperatures (Fig. 6). Probably, the most 
surprising feature is the absolute value of the period of current oscillations $\Delta \Phi_{I}$ in units of 
magnetic flux quantum $\phi_{0}$. The period increases with an increase of the loop size and reaches a value 
$\Delta \Phi_{I} / \phi_{0} \sim 34 \pm 3$ for the largest 25 $\mu$m $\times$ 25 $\mu$m structure (Fig. 7). 
The uncertainty in the definition of the effective loop area, due to the finite line width, cannot account 
for a such high value of discrepancy. Although the majority of experiments were made in the voltage biased 
mode $I(V=const, B)$, the current biased dependencies $V(I=const, B)$ were also measured (Fig. 8). 
Qualitatively, the same oscillating behaviour with hysteresis in the magnetic field was observed. 
However, there are several important differences. Firstly, the shapes of the voltage and the current oscillations 
differ: the $V(I=const, B)$ 'parabolas' are 'upside-down' (Fig. 8a). Secondly, the normalized magnitude of 
the voltage oscillations is much smaller than the corresponding current variation taken at the same point 
of the I-V characteristic (Fig. 8b). Thirdly, the period of the voltage oscillations $\Delta B_{V}$ is smaller 
than the corresponding period of the current oscillations $\Delta B_{I}$  (Fig. 8c).

The authors have no solid explanation for the  mentioned phenomena. The most 
confusing feature is the deviation of the period of oscillations in the magnetic field from the expected 
value $h/e$ or $h/2e$. 
The allowed states of a superconducting ring differ by the phase change accumulated over the circumference of
the loop. The energy of the $n$-th state is given by \cite {Glazman}:
\begin{equation}
\mathcal{E}_{n} = \frac  {2 \pi^{2} \hbar ^{2} n_{s} \sigma}    {m^{\ast} S}    %
(\frac {\Phi} {\phi _{0}^{S}} + n )^{2},
\end{equation}

where $n_{s}$ is the density of superconducting electrons, $m^{\ast}$ is the effective electron mass, $\sigma$ - 
cross section area of the wire, forming the loop, $S$ - the loop's circumference, $\Phi$ is the magnetic flux 
through the area of the loop, and $\phi _{0}^{S}= h/2e$ is the 
superconducting flux quantum. The persistent current is proportional to the derivative of the energy, 
$I \sim dE / d \Phi$, and shows the 
characteristic sawtooth behavior with a period $\Delta\Phi _{I} = \phi _{0}^{S}$. What is measured in the experiment
is the transport current throught the whole N-I-S structure (Fig. 1), and not the persistent current.  However,
any explanation based on purely 'superconducting' considerations should result in a $h/2e$ periodicity 
independent of the size of the system (Fig. 7), range of the magnetic fields (Fig. 2) and the measuring mode 
(voltage or current bias)(Fig.8).

In present measurements three different coils were used, each being calibrated at room temperature and at 4.2 K. The 
data was found to be quantitatively consistent. The coils were wound directly on the metal vacuum shield of the 
$^{3}He-^{4}He$ dilution refrigerator. The canister material (welded stainless steel) and the sample chamber (copper) 
contained no superconducting alloys, which might have attenuated the magnetic field. Nevertheless, a control test 
was made. A pure aluminium 5 $\mu$m $\times$ 5 $\mu$m loop which contained no other materials and 
no tunnel junctions was fabricated. Oscillations of the sample's conductivity, while in a resistive state 
(Little-Parks effect) 
\cite {Little-Parks}, were measured using the same experimental set-up. Due to a geometrical pre-factor the magnitude 
of the effect was rather small $\Delta R/R \sim \Delta T_c / T_c \sim (\xi/L)^{2}(n - \Phi / \phi _{0}^{S})^{2}$
($\xi$ being the superconducting coherence length and $L$ is the loop side) 
\cite {Tinkham book}, but still the oscillating behaviour R(B) was reliably characterised. The period of oscillations
was equal to the expected value $\Delta\Phi = \phi _{0}^{S}= h/2e$ within a reasonable accuracy $<$ 5\%.
The data gives the confidence that the observed periodicity in the N-I-S systems is not a product of the measurement 
hardware artefacts. It is important to mention that the deviation of the period of the Little-Parks 
oscillations in mesoscopic superconductors from constant behaviour \cite {Mosch-square} is associated with 
the formation of a superconducting sheath at high magnetic fields of relatively 'bulk' systems 
(essentially not one-dimensional). The deviation of the period of oscillations $\Delta \Phi _{I}$ and 
$\Delta \Phi _{V}$ from $h/2e$ in N-I-S quasi-one-dimensional structures (particularly, those with the 
loop side $L$ larger than a few $\mu$m) can by no means be explained by the mentioned effect \cite {Mosch-square}.

A possible alternative explanation is a multiple vortex penetration within the superconducting 
'walls' of the structures while in a mixed state. However, the effective core size of a single vortex is about 
the dirty-limit 
coherence length $\xi \sim (\ell \xi _0)^{1/2}\sim$ 150 nm ($\ell$ being the mean free path and $\xi _0$ is BCS
coherence length), and is not much smaller than the line width of the studied N-I-S structures. Thus, there is 
not enough room for vortices to fit within the 'walls' of the superconducting loop. 
The period of oscillations originating from the possible vortex penetration within the Cu-AlO-Al overlapping area 
(Fig. 1) is too small compared to the experimantally observed. It should not depend on the 
loop size, being randomly varying for each particular sample close to nominal $\sim$ 100 nm x 100 nm. 
Additionally, the penetration of a magnetic vortex inside a type-II superconductor requires the overcoming of 
a temperature-dependent potential barrier. The latter results in non-monotonous, strongly hysteretic, 
random (non-periodic) magnetic field patterns, which are essentially temperature dependent. 
The observed oscillations are reasonably periodic, well reproducible and the periods $\Delta \Phi _{I}$ and 
$\Delta \Phi _{V}$ are nearly temperature independent. The test 
structure which consisted of a solid 5 $\mu$m x 5 $\mu$m square overlapped through a tunnel 
barrier by a copper electrode was studied. The I-V characteristic was of a conventional type for a N-I-S system
(Fig. 2), while the variation of the current in a magnetic field $I(V=const, B)$ showed complicated 
non-monotonous behaviour with no signs of periodicity. This behaviour agreed with the expectations, mentioned above.  
Thus, the origin of oscillations due to multiple vortex penetration in the studied N-I-S 
quasi-one-dimensional systems  should be ruled out.

It has been proposed \cite {Esteve private} that oscillating behaviour in our geometry might originate 
from the modification of the energy gap and the density of states by the screening (and transport) currents 
when the flux threads the loop. Certainly, such effect should contribute to some extent, but a 
noticeable impact is only expected at high magnetic fields (currents) comparable to 
the critical ones \cite {Sanchez}. More importantly, the periodicity should be equal to the usual  
'superconducting' value $h/2e$.

Most likely, similar oscillating behaviour have been reported for a single-electron-transistor (SET) 
composed of superconducting 
central island in a form of a loop \cite {Jap SET}. The period of oscillations was found to be sometimes 
higher than $h/2e$. Furthermore, it was not constant in a magnetic field and was different for voltage and 
current biased modes. 
Unfortunately, no solid explanation applicable to our geometry has been proposed \cite {Jap SET}.

Several possibile explanations of the reported phenomena have been outlined above. All of them are related
to 'purely superconducting' properties.  
Experiments involving N-I-S tunnel junctions allow one to pump nonequilibrium quasiparticles from
a normal electrode into a superconductor. These nonequilibrium excitations have finite lifetimes (relaxation 
lengths) and may interfere at shorter scales. 
Relaxation of nonequilibrium quasiparticles inside a superconductor has been studied 
intensively over 20 years ago \cite {Clarke in Gray}, \cite {Tidecks}. Formation of a Cooper pair from 
injected quasiparticles with energy $E$ is governed at least by three processes. The corresponding 
lifetimes are: quasiparticle scattering ($\tau _S$), branch imbalance ($\tau _Q$) and recombination ($\tau _R$). 
The first process is responsible for quasiparticle scattering with the emission/absorption of a phonon. 
The second process equalises the population of electron-like and hole-like excitations. The third process 
governs the recombination of the two quasiparticles above the gap into a Cooper pair at the Fermi level.
It has been shown theoretically that all the mentioned lifetimes 
tend to infinity at energies $E$ of the order of the superconducting gap $\Delta$, 
and all decrease rapidly at higher excitations \cite{Kaplan}. The majority of experiments at that time were 
performed close to a superconducting critical temperature. The low temperature limit was studied very poorly. 
However, some data can still be found. For example, the charge imbalance relaxation 
time for aluminium at $T \ll T_{c}$ was measured, and it can easily exceed $\sim$ 10 ns \cite{Clarke in Gray}. 
To the knowledge of the authors, 
the subject of nonequilibrium quasiparticle interference has never been addressed neither theoretically, nor 
experimentally. If a reasonable assumption is made that there are no other inelastic mechanisms 
involved, the quasiparticle phase breaking time should be equal to the smallest of the lifetimes mentioned above :
$\tau _{\varphi}^{Q} = min(\tau _S, \tau _R, \tau _Q)$. 
Following \cite {Kaplan} one comes to a conclusion that, at injection energies $eV \sim \Delta$ , the
nonequilibrium quasiparticles can preserve their phase on macroscopic distances 
$\ell _{\varphi}^{Q} \sim (\tau _{\varphi}^{Q} D)^{1/2}$, where $D$ is the diffusion coefficient 
(about 100 $cm^{2} / s$ for studied aluminium films). 

By associating the observed oscillations with the interference of nonequilibrium quasiparticles, one can qualitatively 
explain some features of the phenomena. The positions of the maxima $\Delta I/I_{max} (V/V_{gap})$, which appear 
close to the gap energy $\Delta= eV_{gap}$ (Fig. 5), is related to the maximum of the energy dependence of the  
quasiparticle density of states exactly at $E=\Delta$. Thus, at energies $E \sim \Delta$ more quasiparticles 
can participate
in interference. The corresponding temperature behaviour (Fig. 5) reflects the 'sharpening' of the 
quasiparticle density of states energy dependence with the decrease of the temperature.
However, there are several serious objections against such a proposal. 
Firstly, the oscillations do not decay rapidly at excitations below the energy gap (Fig. 5). 
Secondly, as the current on I-V characteristics is still finite at $eV < \Delta$ (Fig. 2), this means that either 
quasiparticles 
or Cooper pairs do transport the current. One might expect a doubling of the period of oscillations 
from $h/e$ to $h/2e$
while sweeping the bias voltage through the gap value $V_{gap}$. However, nothing definite happens when this is done
(Fig. 6). 
Thirdly, 5$\mu$m $\times$ 5$\mu$m S-I-S structure (all aluminium ) showed the oscillating behaviour. 
The shape of the $I(V=const, B)$ dependencies apeared to be different from the Cu-AlO-Al system, but the period was 
close to the N-I-S case. Finally, the last problem is the absolute value of the oscillation period. 
Assuming that the periodicity $\Delta \Phi = h/q$ originates from the interference of quasiparticles, one should 
require that these nonequilibrium excitations have fractional charge $q < e$. 
Mathematical treatment has been described \cite {Tinkham book}. The effective charge of a quasiparticle 
changes continuously from 
electron-like $q_k=(u_{k}^{2} - v_{k}^{2}) = \xi _{k}/E_{k} \approx 1$ to hole-like $q_{k} \approx -1$ as one goes from 
outside to inside the Fermi surface.
The effective charge $q_k$ is much smaller than unity as $E_{k}$ approaches the energy gap $\Delta$. 
Here $E_{k} = (\Delta^{2}+ \xi _{k}^{2})^{1/2}$, $\xi _{k}$ is the one-electron energy of state $k$, and
$v_k$, $u_k$ are BCS occupation probabilities. Unfortunately, it looks very doubtful if this mathematical 
formalism represents the 'real' charge, which is responsible for the electric current in N-I-S systems. 
Probably, shot noise experiments are required to answer this question. Additionally, it is not 
clear why for the wide range of injection energies $eV$ the charge becomes 'more fractional' as the size of 
the loop increases. The origin of the difference of periods in current and voltage biased modes (Fig. 8) 
is by no means clear. Probably, an interplay between the transport current, set by the external source, and 
the screening supercurrent is important for the explanation.

In summary, an unusual oscillating phenomenon has been observed in N-I-S non-single-connected microstructures (Fig.1). 
The results of measurements on various structures with differnet geometry are quite consistent. Probably, the most 
surprising is the period of oscillations in units of magnetic flux, which noticeable exceeds the value $h/e$ 
for larger loops (Fig. 7). The authors have no solid explanation of the observed phenomena. Further experiments 
and theoretical analysis are required.

The authors would like to acknowledge D. Esteve and J. Pekola for their helpful discussions, and D. Agar for the
help with the manuscript. The work was supported by the Russian Foundation for Basic Research (Grant 01-02-17427) 
and the Academy of Finland under the Finnish Centre of Excellence Program  2000-2005 No. 44875, 
Nuclear and Condensed Matter Program at JYFL.

%%%%%%%%%%% FIGURES %%%%%%%%%%%%%%%%%%%%%%%%%%%%
\clearpage

\begin{figure}
\caption{Scanning electron microscope image of a structure with 3 $\mu$m $\times$ 3 $\mu$m loop . Schematics of  
voltage-biased measurements.}
\label{Fig. 1}
\end{figure}

\begin{figure}
\caption{Sample REFR-61-C@09.04.02 with 5 $\mu$m $\times$ 5 $\mu$m loop. a) Typical $I(V)$ (left axis) and 
$dI/dV(V)$ (right axis) characteristics at zero magnetic field $B$. 
Symbols $\pm V_{gap}$ indicate positions of the gap voltage. Symbols with arrows show the bias voltages, 
where the magnetic field
sweeps in Fig. 4 were taken. b) Zooms of the same I(V) characteristic at various magnetic fileds.}
\label{Fig. 2}
\end{figure}

\begin{figure}
\caption{Sample REFR-91-B@02.08.02 with 25 $\mu$m $\times$ 25 $\mu$m loop. $I(V/V_{gap}= 0.31,B)$ dependence 
at $T= 218 \pm 3 $ mK. The arrow indicate the direction of the field sweep. Inset: corresponding magnetic 
field dependence of the oscillation period $\Delta B_{I}$.}
\label{Fig. 3}
\end{figure}

\begin{figure}
\caption{Sample REFR-61-C@09.04.02 with 5 $\mu$m $\times$ 5 $\mu$m loop. $I(V=const,B)$ dependencies taken at
 various bias points $V / V_{gap}$ , $T= 165 \pm 5 $ mK. Symbols correspond to notations in Fig. 2. The solid-line 
 arrow shows direction of the field sweeps for the solid symols, the dashed line - for the open symbols. }
\label{Fig. 4}
\end{figure}

\begin{figure}
\caption{Normalized magnitudes of the low field ($\mid B \mid < 2 $ mT ) current oscillations 
$\Delta I/I_{max}$ as a function of the 
normalized bias $V/V_{gap}$ for three different samples at various temperatures.}
\label{Fig. 5}
\end{figure}

\begin{figure}
\caption{Sample REFR-71-B@16.05.02 with 10 $\mu$m $\times$ 10 $\mu$m loop. Dependencies of the period of the 
current oscillations $\Delta B_{I}$ measured at low magnetic fields ($\mid B \mid < 2 $ mT )
 on the normalized bias $V/V_{gap}$ for two temperatures.}
\label{Fig. 6}
\end{figure}

\begin{figure}
\caption{The period of the current oscillations $\Delta \Phi _{I}$ in units of the magnetic flux quantum
$\phi _{0}^{N}= h/e$ measured at low magnetic fields ($\mid B \mid < 2 $ mT )  
(left axis, circles), and the maximum magnitude of the normalized current amplitude $\Delta I / I_{max}$ 
(right axis, triangles) at temperatures $T<100$ mK as functions the loop circumference $S$.
The dotted line is a guide for the eye.}
\label{Fig. 7}
\end{figure}

\begin{figure}
\caption{Sample REFR-63-B@29.03.02 with 5 $\mu$m $\times$ 5 $\mu$m loop.
a)	Current  and voltage oscillations, corresponding to 
the same point of the I-V characteristic at zero magnetic field. $V / V_{gap} =$ 1.17, $T =$ 127 $\pm$ 1 mK. 
The dotted lines are guides for the eye.
b)	Normalized magnitudes of the current and voltage oscillations as  
functions of the normalized bias voltage $V / V_{gap}$ (or $V(B=0)/V_{gap}$ in case of the current-biased mode).
c)	Periods of the current and voltage oscillations as functions of the same argument as in b).
For the voltage-biased data the solid symbols and the left axis are used, for the current-biased - the 
open symbols and the right axis.
For figures b) and c) only data at low magnetic field ($\mid B\mid < $ 2 mT) was considered.}
\label{Fig. 8}
\end{figure}

%%%%%%%%%%% BIBLIOGRAPHY %%%%%%%%%%%%%%%%%%%%%%%
\clearpage

$^\ast$ Also at Moscow State University, Department of Physics, Institute of Nuclear Physics, Moscow 119899, Russia.

\end{document}